# DNA NANOROBOTICS

*Mustapha Hamdi and Antoine Ferreira*

Laboratoire Vision et Robotique, ENSI Bourges - Université d'Orléans, 18000, Bourges, France.
{mustapha.hamdi, antoine.ferreira}@ensi-bourges.fr

**ABSTRACT**

This paper presents a molecular mechanics study of nanorobotic structures using molecular dynamics (MD) simulations coupled to virtual reality (VR) techniques. The operator can design and characterize through molecular dynamics simulation the behavior of bionanorobotic components and structures through 3-D visualization. The main novelty of the proposed simulations is based on the mechanical characterization of passive/active robotic devices based on double stranded DNA molecules. Their use as new DNA-based nanojoint and nanotweezer are simulated and results are discussed.

**1. INTRODUCTION**

There has been a great interest and many reports in the use of DNA specifically to actuate and assemble micro- and nano-sized systems. In nano-electronics, DNA could be used as molecular switchs for molecular memories or electronic circuitry to assemble future nano-electronics transistors [1],[2]. In nanorobotics (Fig.1), structural elements could be carbon nanotubes while the passive/active joints are formed by appropriately designed DNA elements [3]. In such nanodevices, nature assembles nano-scene components using molecular recognition [4]. In the case of DNA, hydrogen bonding provides the specificity behind the matching of complementay pairs of single-stranded (ss) DNA to hybridize into a double strand (ds) of helical DNA. While these tasks have been performed by nature efficiently and perfectly, roboticians and engineers should use prototyping tools for a CAD design process of future bionanorobotic systems. To achieve these long-term goals, prototyping tools based on molecular dynamic (MD) simulators should be developed in order to understand the molecular mechanics of proteins and develop dynamic and kinematic models to study their performances and control aspects. The ability to visualize the atom-to-atom interaction in real-time and see the results in a fully immersive 3-D environment is an additional feature of such simulations [5]. Virtual Reality (VR) technology is applied here, which not only provides immersive visualization but also gives an added functionality of CAD-based design, simulation, navigation and interactive manipulation of molecular graphical objects. Using simulated biological nano-enviroments in virtual environments, the operator can design and characterize through physical simulation the behavior of molecular robots. Adding haptic interaction, the operator can explore and prevent the problems of bionanorobots in their native environment.

Several types of software for displaying 3D models of bio-molecules have been developed. Most of these programs provide haptic manipulation and 3D visualization in combination with molecular dynamics simulation such as Visual Molecular Dynamics (VMD) [6], that can be connected to Molecular Dynamics Simulation program (NAMD) [7] via Interactive Molecular Dynamics Simulation (IMD) [8] and Virtual-Reality Peripheral Network (VRPN) [9] (the reader can refer to [10] for a good introduction to the domain).

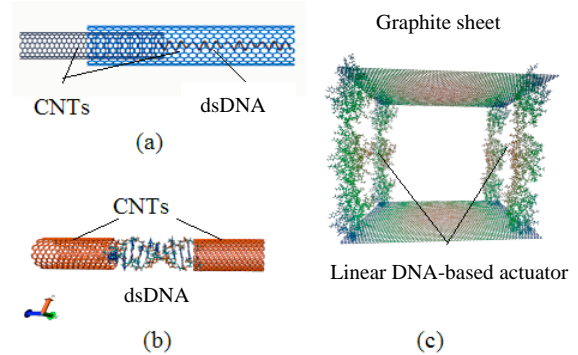

**Fig.1**: Basic DNA-based molecular components. Passive joints (a) 1 d.o.f. and (b) multidegree of freedom elastic spring. Their elastic behavior can be used as a passive control element or as the restitution force that will bring the mobile component back to its original position. Active joints (c) single degree of freedom parallel platform. The platform is actuated by double stranded DNA-type linear nanoactuator.

Based on VR technology and MD simulators, our long-term goal is to prototype virtually bionanorobotic systems and optimize mechanically their design considering the native microenvironment. In this work, we consider real-time force-feedback technology for improving the methodology of mechanical molecular studies of DNA molecules acting as passive and/or active joints in bionanorobotic systems, as illustrated in Fig.1. Understanding the molecular mechanics behavior of such DNA molecules is an important challenge for interfacing DNA to structural (CNTs, graphite) or biological (biological nanotube, proteins) elements.

The paper is organized as follows. First, section II presents the developed virtual CAD simulator Then, different design concepts of DNA-based devices are proposed: section III deals with DNA-based springs and section IV introduces a novel DNA-based nanogripper.

## 2. CAD SYSTEM BIO-NANOROBOTIC PROTOTYPING

The application of advanced virtual reality techniques in bio-nanotechnology allowed to study, simulate, visualize and interact with molecular biology, genomics, proteomics, structural and computational biology. As far we know, only basic molecular elements have been studied until now. As a perspective of these developments, VR-based technology could be used for the molecular robot design. Several types of software for displaying 3D models of bio-molecules have been developed supporting VR devices such as stereo glasses, 3D trackers, and force-feedback (haptic) devices. Real-time exploration is what attracts researchers to develop man-machine interfaces for nanoscale manipulation that use haptic display technologies. The developed simulation system presented in Fig.2 permits manipulation of bionanorobotic components during molecular dynamic (MD) simulations with real-time force feedback and 3D graphical display. It consists of three primary components: a haptic device controlled by a computer that generates the force environment, a MD simulation for determining the effects of force application, and a visualization program for display of the results. Communication is achieved through IMD protocol between the visualization program Visual Molecular Dynamics (VMD) and the molecular dynamics program (NAMD) running on multiple machines. A force-feedback PHANToM device measures a user's hand position and exerts a precisely controlled force on the hand in order to apply different mechanical constraints, force and energy fields on the virtual model in order to prototype bionanorobotic components. The application of the mechanical constraints are applied by a virtual AFM microlever and can be manipulated with a mouse or a haptic interface with force-feedback. The principal idea of this application is to let the user drive le molecule in the space of precomputed conformations and transitions:

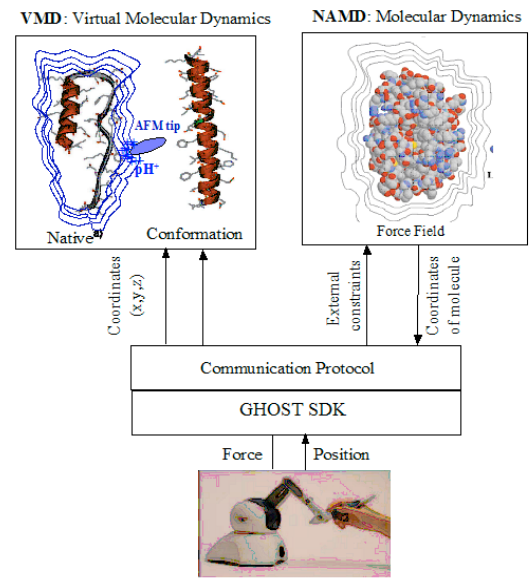

**Fig.2.** Basic concept of virtual environment and haptics technology for bionanorobot simulation.

- The user steers virtually he molecule protein with a cursor, which is a small spherical ball that is position controlled by the stylus. The system allows application of stretching, shearing and bending mechanical constraints on the protein.
- The user distinguish possible conformational paths and feels the energy barriers associated with particular paths.

## 3. DESIGN OF CONTROLLABLE DNA PASSIVE SPRINGS

### 3.1. Mechanical Characteristics of dsDNA

Chemically, DNA is a long polymer made up of a linear series of subunits known as nucleotides. Structurally, DNA is usually found as a double helix, with two strands wrapped around one another. However, DNA can adopt other configurations and can also exist in single-stranded forms. Double-stranded DNA (ds-DNA) has sparked the renewed interest in the force versus extension of polymers for biomolecular springs. The DNA is solvated in water with 30 $Na^+$ ions added to neutralize the charge. The water-DNA system was gradually heated over 7ps to 300K, and then equilibrated with a thermal bath at 300K for another 7ps. The dynamics of the DNA as shown in Fig.5 were performed with the double-stranded terminations stretched and the other end fixed. This elastic behavior is thus purely entropic. For very low tension $f \leq 1pN$, the restoring force is provided by "entropic elasticity".

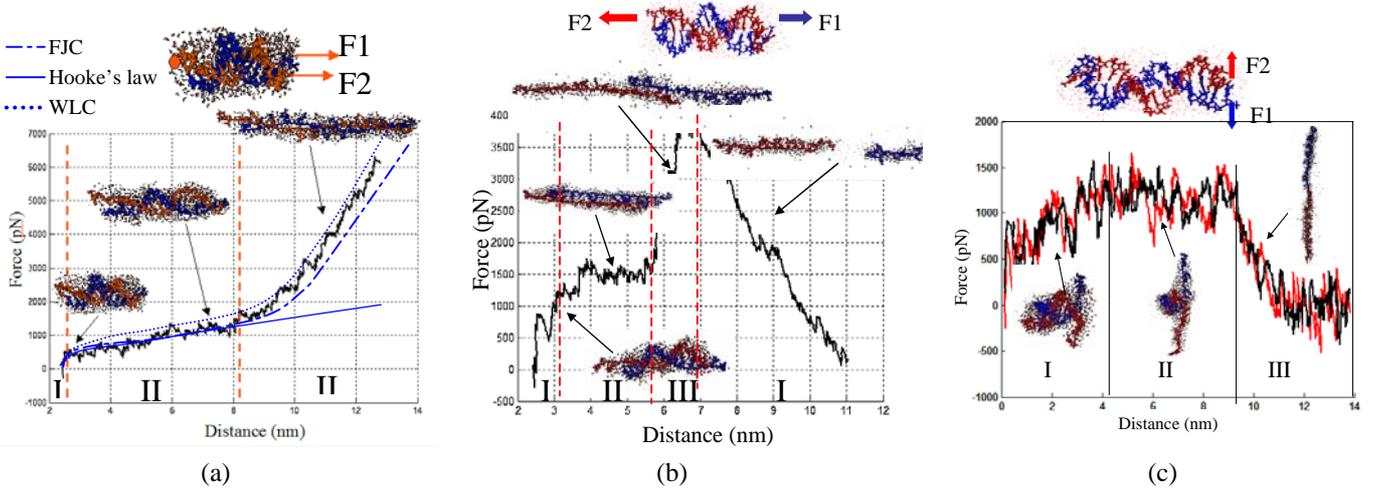

**Fig.5:** Double-stranded DNA mechanical simulations: (a) force-extension curve when applying a stretching force. The data are fit to a WLC model solved numerically assuming *A*=43nm. The FJC curve assumes *b*=2*P*=106nm. (b) force-extension curve when applying an anti-parallel stretching and (c) force-extension curve when applying an anti-parallel shearing.

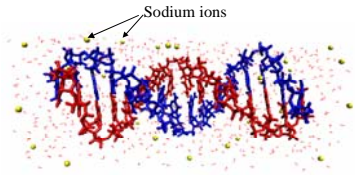

**Fig.4:** Structure of water-DNA system ionized with 3O Na$^+$ ions.

In the absence of any force applied to its ends, the DNA's RMS end-to-end distance (chain length, *L*) is small compared to its contour length defined as the maximum end-to-end distance (maximum length, $L_0$) and the chain enjoys a large degree of conformational disorder. Stretching DNA reduces its entropy and increases the free energy. The corresponding force *f* increases linearly as a Hooke's law with the extension *L*:

$$f \cong \frac{3k_B T}{A_{DNA}} \frac{L}{L_0}, \qquad L \ll L_0 \qquad (1)$$

The length $A_{DNA}$ is known as the "thermal persistence length" of DNA and is of the order 50nm. For higher forces ($f \geq 10$pN), the end-to-end distance *L* is close to $L_0$ and the elastic restoring force is due to distortion of the internal structure of DNA. In this regime, the force extension curve can be approximated by two models that are often used to describe the entropic elasticity of DNA. In the freely jointed chain (FJC) model, the molecule is made up of rigid, orientationally independent Kuhn segments whose length, *b*, is a measure of chain stiffness. The resulting entropic elastic behavior can be summarized in the force-extension relation :

$$\left\langle \frac{z}{L_{tot}} \right\rangle = \coth\left(\frac{f b}{k_B T}\right) - \frac{k_B T}{f b} \qquad (2)$$

defining the well-known Langevin function. Expanding Eq.(2) gives the effective spring constant for the FJC as $k^{FJC} = 3k_B T / b$. The terms $L_{tot}$ represent the total length of the protein and *f* the stretching force. The alignment of segments by tension is described by Boltzmann distribution. In the inextensible worm-like chain model (WLC) model [11], the molecule is treated as a flexible rod of length *L* that curves smoothly as a result of thermal fluctuation. The WLC model of entropic elasticity predicts the relationship between the relative extension of a polymer (*z/L*) and the entropic restoring force (*f*) through

$$f = (k_B T / A)\left[z / L + (1/4(1 - z/L)^2) - 1/4\right] \qquad (3)$$

where $k_B$ is the Boltzmann constant, *T* is absolute temperature, *A* is the persistence length, *z* is the end-to-end length, and *L* is the length. Results, shown in Fig.5a, indicate that, even though the FJC model can describe the behavior of ds-DNA in the limit of low and intermediate forces, it fails at high forces. The WLC model, on the other hand, provides an excellent description of molecular elasticity at intermediate and high forces. Both models behave as a Hooke's law for low stretching forces. In the case of an anti-parallel stretching (Fig.5b), the force-extension curve is quite similar for the Parts I and II when the ds-DNA is unfolding. When it is totally stretched, lateral mechanical unzipping of the hairpins occurs sequentially leading to an increase of the stretching force (Part III) until to reach its point of rupture (Part IV).

When considering an anti-parallel shearing of ds-DNA (Fig.5c), a mechanical longitudinal unzipping of the 12 hairpin DNA occurs sequentially (Part I) after a phase of extension of ds-DNA structure (Part II) before the rupture point (Part III). As it is shown, the strand unzipping occurred abruptly at 500pN and displayed a reproducible "saw-tooth" force variation pattern with an amplitude of +/-10pN along the DNA.

### 3.2. Passive DNA-SWNT Elastic Joint

Considering the interesting stretching characteristics of ds-DNA of (Fig.5a), we designed a passive DNA-based elastic joint (see Fig.6a) composed of two single walled carbon nanotubes (SWNT) as structural components. The covalent linkage between DNA-SWNT are realized at the terminations through atoms of carbon (Fig.6b). As shown in Fig.7, the passive dsDNA-based joint is fully reversible in the domain of linear elasticity of the DNA molecule (Part I and II). The spring constant is estimated to be as $K_{DNA}$= 780 N/m with a thermal bath at 300K and constant acidic pH=5.

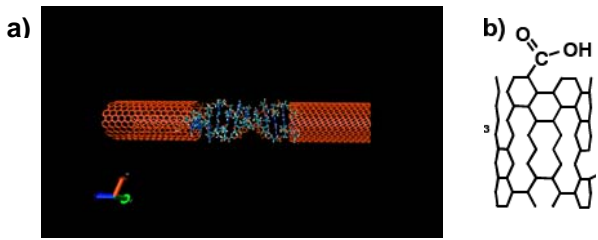

**Fig.6:** (a) Passive joint composed of a dsDNA protein attached at both ends by carbon nanotubes (CNT). The water molecules are not shown. (b) schematic illustration of chemical scheme for SWNT-DNA linkage.

The modulation of the strength of the molecular nanospring has been considered. Variations of the medium environment have been controlled through three parameters: (i) the temperature, (ii) the level of acidic pH and (iii) the electron density of the molecule. In this regard, a molecular variable spring can be envisaged. Fig. 8a shows the evolution of the DNA spring constant $K_{DNA}$ with respect to the temperature ranging from 273 K to 500 K. The results shows clearly the linear decrease of spring constant with respect to the temperature (between 300K to 450K). The adjustable stiffness could be adapted between 780 N/m to 720 N/m in relation to the temperature control. However, for such high temperatures the DNA becomes completely denaturated and reversibility is altered by the strong hysteresis. Practically, the DNA molecule is able to perform repeatable motion controlled by variation of pH by adding protons (termed *protonation*). It is proposed to isolate this domain from the DNA and trigger the stiffness change by variation of pH. The drop of pH changes the energetics of the outer (enveloppe glycoprotein) protein of the DNA in such a way that there is a distinct stretching change in a part of it.

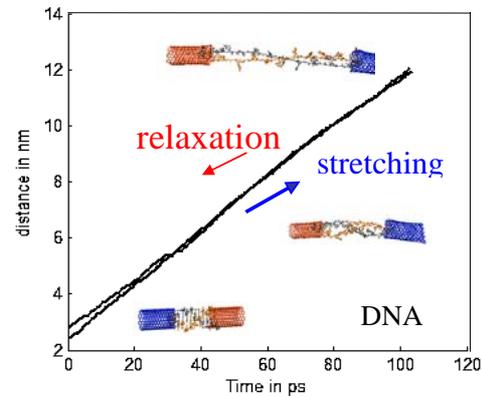

**Fig.7:** Reversibility of the DNA-based elastic joint during a stretching-relaxation phase.

Fig.8b shows the change of the protein stiffness with respect to the pH value. For a neutral acidic pH value (pH=11), the DNA protein is in a partially α-helical stranded coiled coil. The value of pH=5.5 corresponds to the neutral state. An increase occurs at lower pH between 5.5 and 10 leading to a stiffness variation ranging from 670 N/m to 780 N/m. For this configuration, it is necessary to protonate the amino acid side chain of the protein by adding proton inside the native environment. Finally, the increase or decrease of the electron density (namely, the electric charge *Q* in Coulombs) has been investigated. The concept of using electrons as a medium for such purposes is not unreasonable because electron density will readjust along the molecular backbone in response to an inflow/outflow of electronic charge. Moreover, using electrons as a mechanism for changing the shape and rigidity of molecular systems has precedence [12]. Fig.8c shows the change of spring stiffness according to the electron charge of the protein. From these results we derive that the force stiffness bracket a range from 680 N/m to 780 N/m for the ds-DNA protein. Experimentally one can add or remove electrons with a potentiostat, or by appending suitably placed electron-donating or withdrawing groups. Single-stranded DNA (ss-DNA) and double-stranded DNA (ds-DNA) proteins have interesting characteristics due to their elastic behavior and structural conformation [13]. As example, DNA can provide three independent reversible degrees of freedom: stretching, shearing and twisting. They can be used to construct molecular building blocks by virtue of their conformation pathways. Branched DNA molecules with sticky ends are promising for self-assembling macromolecular periodic structures [14].

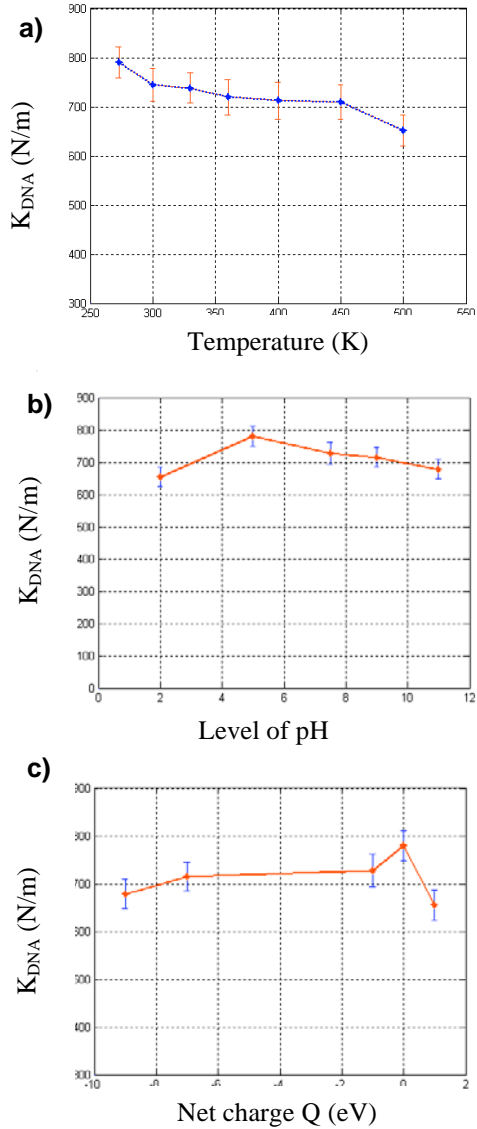

**Fig.8:** Evolution of the spring stiffness of ds-DNA during stretching for changing medium properties: (a) temperature, (b) level of acidic pH and (c) level of electron charge Q.

## 4. CONTROLLABLE DNA-BASED NANOGRIPPER

In this part, we study the molecular properties of DNA proteins to change their 3D conformation depending on the temperature level of the native medium. Thus, a new biomolecular gripper type called dsDNA-SWNT nanogripper is designed (Fig.9). It is composed of a dsDNA protein with two SWNT as nano end-effectors. The structure is like an hairpin composed of two coils, having each C-terminal connected to a SWNT and undergoing a conformational change induced by temperature variation (increase or decrease) in mildly acidic conditions (i.e., pH around 5.5). With the change of temperature, the N-terminals pop out of the inner side and the peptide acquires a straightened position. Two known states are established, i.e., the *native* state and the *fugeonic* state of the 12-hairpin peptide of DNA. The representative "open" structure can be generated by forcing the structure away from the native conformation to open state (conformation) with constrained high-temperature molecular dynamics (molecular dynamic program NAMD with the CHARMM22 force field). Both the forward (closed-to-open) and the reverse (open-to-close) transformations are carried out for two end-point structures of 7Å. The dsDNA is able to perform repeatable motion controlled by variation of temperature. As example, Fig.10 shows the front view trajectory of both free terminations of dsDNA when considering temperature augmentation from 300K to 400K.

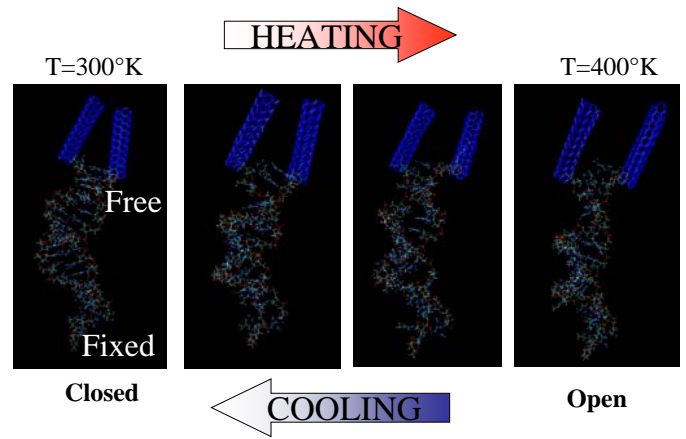

**Fig.9:** dsDNA-SWNT nanogripper. In the native state the nanogripper is closed (T=300K) and in the conformational state (T=400K) the nanogripper is open.

The finger's motion are composed of a linear finger opening (*y-z* plane) connected to carbon nanotubes as nano end-effectors. In order to assess the interaction forces when handling a nanoobject, we simulated a contact force by using two nonlinear spring $K_1$ and $K_2$ at each end-effector. Fig.11 shows the evolution of the gripping force during the opening of the nanotweezers. The contact force is estimated by $F_i=K_i \Delta x_{i,j}$ with $(i,j)=(1,2)$ for each nanofinger. It should be noticed that when the nanofingers are in closed contact with the nanoobject, the forces provided by the ds-DNA are quite linear with the opening displacement $\Delta x$. For large opening displacements, the forces are dissimilar due the small parasitic rotation of the fingers. Fig.12 shows the motion reversibility of the nanofingers during a temperature cycle time (heating and cooling) from 300K to 400K. Although the overall unfolding pathway is independent of temperature cycling, differences are observed from trajectory-to-trajectory during the cooling procedure. The reversibility error is less than 8% in the worst case.

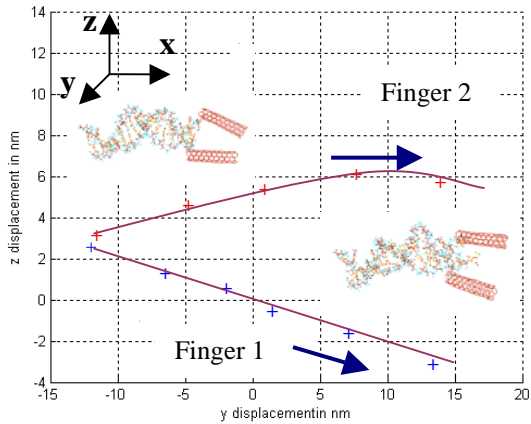

**Fig.10:** Motion of the nanofingers during elevation of the temperature from native state (nanogripper closed) to conformational state (nanogripper open). Influence of the ds-DNA configuration with single walled carbon nanotubes.

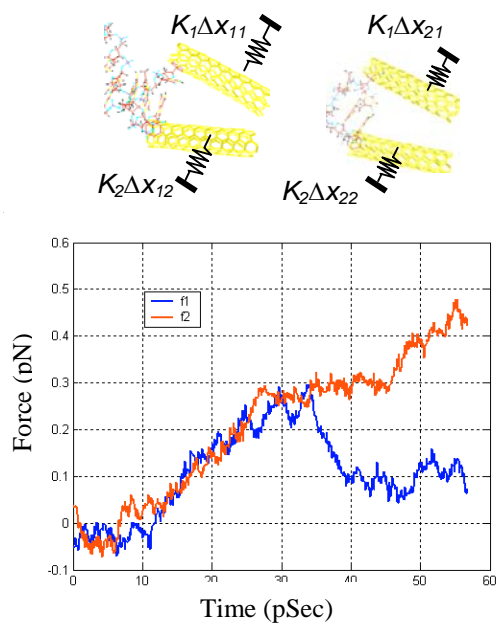

**Fig.11:** Simulation of gripping forces when handling a nanoobject.

## 5. CONCLUSION

This paper presents a computational molecular mechanics study carried out by an interactive CAD simulator interface. The preliminary mechanical force results given in this study corroborates the force results when stretching double-helical DNA molecules for controllable passive elastic joints such as shock absorbers. We demonstrated also that DNA material is well suited to the development of active joints for the control of nanorobotic tasks such as nanotweezers.

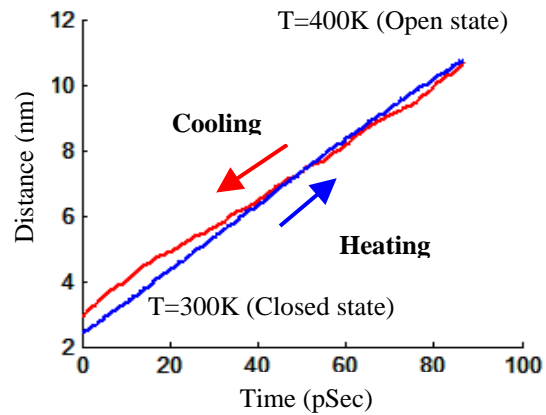

**Fig.12:** Reversibility of the nanotweezers when increasing and decreasing the temperature from T=300K to 400K.